# The computational role of structure in neural activity and connectivity

Srdjan Ostojic[1] and Stefano Fusi[2,3,4,5]

[1] Laboratoire de Neurosciences Cognitives et Computationnelles, INSERM U960, Ecole Normale Superieure - PSL Research University, 75005 Paris, France
[2] Center for Theoretical Neuroscience, Columbia University, New York, NY, USA.
[3] Zuckerman Mind Brain Behavior Institute, Columbia University, New York, NY, USA.
[4] Department of Neuroscience, Columbia University, New York, NY, USA.
[5] Kavli Institute for Brain Science, Columbia University, New York, NY, USA.

**Abstract**
One major challenge of neuroscience is finding interesting structures in a seemingly disorganized neural activity. Often these structures have computational implications that help to understand the functional role of a particular brain area.
Here we outline a unified approach to characterize these structures by inspecting the representational geometry and the modularity properties of the recorded activity, and show that this approach can also reveal structures in connectivity. We start by setting up a general framework for determining geometry and modularity in activity and connectivity and relating these properties with computations performed by the network. We then use this framework to review the types of structure found in recent works on model networks performing three classes of computations.

**Highlights**
- We examine how the structure in neural activity and connectivity is related to the computations a network performs.
- We distinguish two general types of structure that we term geometry and modularity.
- Geometry and modularity can be determined both at the level of neural activity or connectivity.
- We harness these concepts to synthetically review recent modeling works on three classes of computations.

**Glossary**
*Task:* mapping from a set of input stimuli to output actions.
*Latent task variables:* low-dimensional parameters that generate the space of inputs expired in the task.
*Contextual variables:* auxiliary task variables that modify the mapping between stimuli and outputs.

*Neural representation:* mapping from the set of inputs to patterns of neural responses recorded in a brain area or generated in a group of neurons in a network model.
*Activity matrix:* mathematical description of the neural representation in a recording or model network. Each column contains the vector of the neural responses to a particular experimental condition.
*Activity space*: space where each axis represents the activity of one neuron in a recording or model network.
*Selectivity space:* space where each axis represents the selectivity with respect to one task variable.
*Connectivity space:* space where each axis represents an input or output weight of one neuron in a network.
*Geometry:* the spatial arrangement of a set of points in a given space, characterized independently of global rotations or scaling.
*Hyperplane:* generalization of the concept of a plane to a space of arbitrary dimension. A hyperplane splits the space into two halves.
*Dichotomy:* a split of a set of points into two parts, corresponding for instance to two different behavioral outputs.
*Linear separability:* a given dichotomy of a set of points in the activity space is linearly separable if the two parts can be separated by a linear readout or, equivalently a hyperplane in activity space.
*Flexibility of neural representation:* the capacity of a given neural representation to allow a linear readout to implement a large number of distinct dichotomies, or equivalently input-output mappings
*Generalization:* capability to infer correct responses in novel situations, for example the responses to unseen stimuli.
*Abstract representation:* a neural representation that enables generalization with respect to certain variables,
*Disentangled/Factorized Representation:* neural representations that encode information about two task variables along orthogonal directions so that the representation of one variable is invariant with respect to the other variable.
*Modularity:* organization of a set of points in a given abstract space in terms of grouping into clusters, defined based either on their center or shape.
*Mixed Selectivity*: property of individual neurons that respond to combinations of multiple sensory or behavioral variables.
*Functional cell classes:* groups of neurons forming clusters based on their patterns of responses to stimuli. (= in selectivity space?)
*Aligned representations:* neural representations where different groups of neurons encode different task variables.

**Introduction**

In recent years, significant efforts have been deployed to unravel the structure of the brain by establishing detailed atlases of neural cell types based on biological properties such as gene expression, morphology, physiology, or connectivity [1]. The underlying rationale rests on an analogy between neurons and individual building blocks of different kinds, each with a potentially specific function that needs to be understood. However, recordings of neural activity during behavior have revealed a bewildering complexity in the firing of individual cells. A ubiquitous finding is that neurons exhibit *mixed selectivity*, meaning they typically respond to random-looking mixtures of behavioral variables. While initially reported in higher-order areas[2–4], mixed-selectivity has been found across the brain [5–10], concurrently with the observation that both sensory and behavioral variables are represented more broadly across the cortex than previously hypothesized [11–14]. Neural recordings at increasingly large scales have therefore challenged the notion that individual neurons play the role of functional parts with clearly interpretable roles, and raise the question of what type of structure underpins computations that underlie behavior and cognition.

Complex activity and mixed selectivity at the level of individual neurons do not preclude the existence of structure but instead underscore the need to characterize more finely the large spectrum that exists between full randomness and perfect order. Recent works have focused on two types of structure in neural activity: structure at the level of the *geometry* of population representations and dynamics [15–20], and structure at the level of functional categories of neurons[4,21–23], which we refer to here broadly as *modularity* as each category of cells could be considered as a separate module. A key challenge has been to identify the computational implications of different types of structures for behavioral tasks. Neural network models trained using algorithms developed in artificial intelligence have emerged as essential tools to address this question [24–28]. Such networks provide ideal model systems that can learn to perform the same cognitive tasks as animals and humans but are fully observable. Indeed such models provide us with access to the activity of the full network and the underlying connectivity, a crucial additional level of structure that determines both activity and computations.

Here we review recent studies of trained network models that illustrate how the set of computations that a network performs is related to the structure in neural activity and in underlying connectivity. To this end, we start by setting up a framework for characterizing the computational structure in commonly studied behavioral tasks. We next relate this computational framework with characterizations of structure in neural activity based on complementary perspectives of geometry and modularity. We then show how the same approach can reveal geometric and modular structures in the underlying connectivity. We finally use this framework to review the relationship between structure in tasks, activity, and connectivity within three classes of recently studied computations. Altogether, we propose a potential roadmap for interpreting relations between different types of structures present both in behavioral and biological data and in broader classes of artificial neural networks trained on more complex tasks.

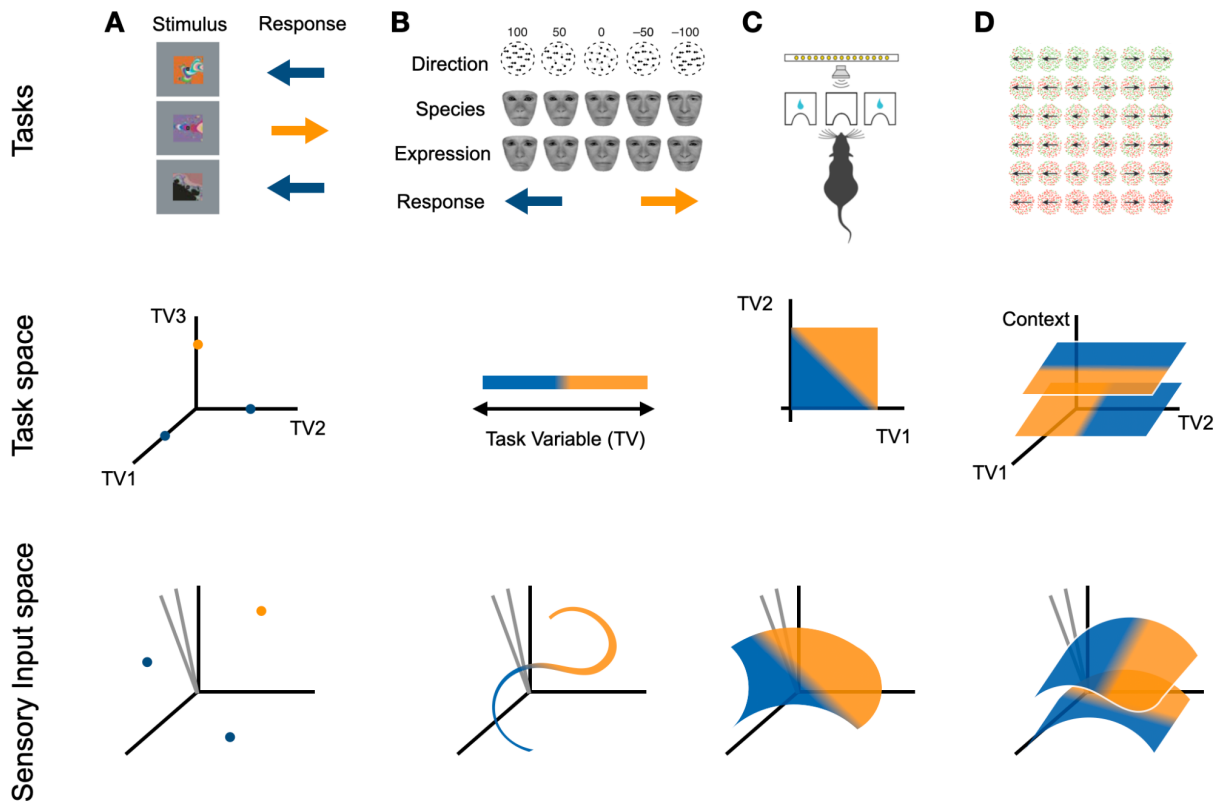

**Figure 1. Characterizing task structure.** Top: schematic illustration of stimuli and responses in four example tasks. Middle: representation of input-output associations in the space of task variables (TVs), where each axis corresponds to a variable controlled by the experimenter. Stimuli are shown as points or manifolds in that space; colors indicate the required responses. Bottom: representation of input-output associations in the sensory input space, where each axis corresponds to the activity of a neuron encoding the sensory input. The sets of responses form non-linear manifolds, where task variables play the role of latent dimensions. A: Stimulus-response association with fractal images [29–31]. B: classification tasks where one continuous task variable defines a morphing between two categories [32]. C: Multi-sensory integration task where the decision needs to be taken by combining two continuous stimuli, e.g., one auditory (TV1) and one visual (TV2) [4]. D: Context-dependent decision-making task where a contextual cue determines which of two continuous stimulus features need to be integrated [3].

1. **Characterizing the computational structure in behavioral tasks**

Following the long tradition of psychophysics, one of the dominant paradigms in systems neuroscience has been to train subjects on simplified tasks partitioned in a series of trials. In each trial, the subject is shown one, or a sequence of, stimuli generated from underlying task variables controlled by the experimentalist. Based on these inputs, the subject needs to produce an action chosen from a typically small set of available options. A task can therefore be formalized as a mapping from a set of inputs, represented as points in the abstract space of task variables, onto required behavioral response (Fig 1 A-B). In this framework, learning a task is equivalent to learning a classification boundary in the space of task variables. A key challenge is, however, that the relationship between the task variables and sensory inputs, such as patterns of activity in the retina, is highly non-linear (Fig 1 B) [33,34]. At the level of neural activity, task variables therefore play the role of *latent variables,* and the intrinsic structure of the task determines how the brain needs to *recode* incoming representations to produce relevant behavioral outputs.

Experimental studies have considered tasks relying on different types of structures. On one extreme, many classical works relied on unfamiliar and almost unstructured stimuli, such as fractal images [29–31]. Each stimulus then defines an independent task condition in a high-dimensional space of task variables (Fig. 1A). On the other extreme, studies on perceptual decision-making and categorization instead focus on more structured stimuli varying continuously along one or several dimensions that define task variables [35–37]. Two typical examples of these variables are the coherence of patterns of randomly moving dots [35] or the variable controlling the morphing from monkey to human faces [32]. In such situations, the latent space of the task variables is low-dimensional, and the required responses vary continuously (Fig.1B), but the sensory inputs are embedded non-linearly in a higher-dimensional space (Fig.1C). In more complex tasks, the required response may be indicated by sequences of stimuli [38–40], require temporal integration [41–43] or depend on additional explicit or implicit contextual variables[3,10,44–47]. Such additional stimuli and context cues increase the dimensionality of the space of task variables (Fig 1C), and the boundaries between different desired responses become more complex (Fig 1D).

## 2. Characterizing the structure in neural activity

How can the structure in the recorded neural activity be characterized and related to the structure of the underlying task? Historically, this question has been pursued using two approaches that focus either on individual neurons or on the population as a whole. Here we review a unifying description that clarifies that these two approaches provide complementary perspectives on the same set of neural activity patterns [18].

Suppose we have access to the activity of a population of N neurons in C trial conditions corresponding to combinations of K task variables. For simplicity, we focus on trial-averaged activity and leave aside trial variability. This dataset forms a CxN *activity matrix* where each row describes the activity of one neuron in the C conditions, and each column stands for the activity of the whole population of N neurons in one condition (Fig. 2). The structure of the neural activity during behavior can then be characterized by studying either the set of columns or the set of rows of this matrix, two approaches that directly correspond to population and single-neuron analyses [18].

Each column of the activity matrix defines a point in the N-dimensional *activity state space*, where each axis is the activity of one neuron [20,48–51]. The resulting description of population activity across conditions is homologous to the representation we used for defining tasks (Fig. 1). In both cases, each task condition is shown as a point colored according to the desired output, but the spaces within which the points live and their overall geometrical arrangement are different. Recent works have exploited a range of metrics based on topology [52–58], distances [17,59–63], or dimensionality [15,50,51,58] to characterize and compare across brain areas the resulting geometry of neural activity. Alternatively, it is instructive to study whether and how a linear readout could map the responses in each task condition onto the output defined by the underlying task [2]. Geometrically, this is equivalent to looking for hyperplanes in the activity state space that separate task conditions according to the desired outputs. Using the simple perspective of a linear readout, an important hypothesis posits that one of the goals of the brain is to transform the sensory inputs to achieve representations that are linearly separable according to the task outputs. Beyond mere linear separability, the analysis of linear readouts across conditions provides a tool for examining to which extent a given neural representation allows for *flexibility*, i.e., the capacity to produce different types of outputs based on a given set of inputs [2,64–66]), or *generalization* by inferring relevant outputs from a subset of task conditions and abstracting away irrelevant features [10,45,67] (Fig. 2B).

A complementary characterization of neural activity is obtained by examining the rows rather than the columns of the activation matrix, thereby focusing on the responses of individual neurons across trial conditions. Classical works have sought to identify *functional classes* of neurons that respond to individual task variables. While this approach has led to important insights, in particular in the primary sensory areas and the navigation system [23], it has become increasingly apparent that individual neurons, in general, exhibit mixed selectivity [2–4], meaning that they respond to mixtures of task variables rather than individual ones. Even if neurons display mixed selectivity, it is still legitimate and interesting to ask whether they could be organized into more general functional classes corresponding to groups of cells with similar responses to multiple task variables [4,21,68,69]. To identify such groups, individual neurons can be represented as points in *condition space*, where each axis is the activity in one of C task conditions [21,68], or in *selectivity space*, where each axis represents some measure of the selectivity to one of the K task variables, for example linear regression coefficients [3,4][4]. The structure in single neuron responses can then be characterized by comparing the resulting cloud of points to a null distribution corresponding to a single isotropic cloud [4,21]. In case of a significant deviation, a clustering analysis can be applied to define groups of neurons sharing particular selectivity patterns. Such analyses uncover an additional level of structure which we denote as *modularity,* that spans the continuum between fully random and pure selectivity (Fig. 2C)*.* The identified groups might correspond to different types of cells (biological modularity), to cells that are connected preferentially to a specific brain area, or simply cells that belong to a certain brain area (anatomical modularity). But different modules might also emerge from learning processes. The significance of this type of modular structure with respect to the underlying computations or biology is only beginning to be uncovered.

Geometry and modularity are two related but distinct views of neural activity at the population level. Modularity implies the existence of functional groups of neurons whose activity spans specific subspaces *aligned* to subsets of the coordinate axes of the activity state space. Instead, geometric analyses in general, focus on properties invariant under rotations in the activity state space. Recent works have argued that such an alignment may play a specific computational role [70–72], particularly when including additional biological constraints such as non-negative activity combined with the minimization of metabolic costs [73].

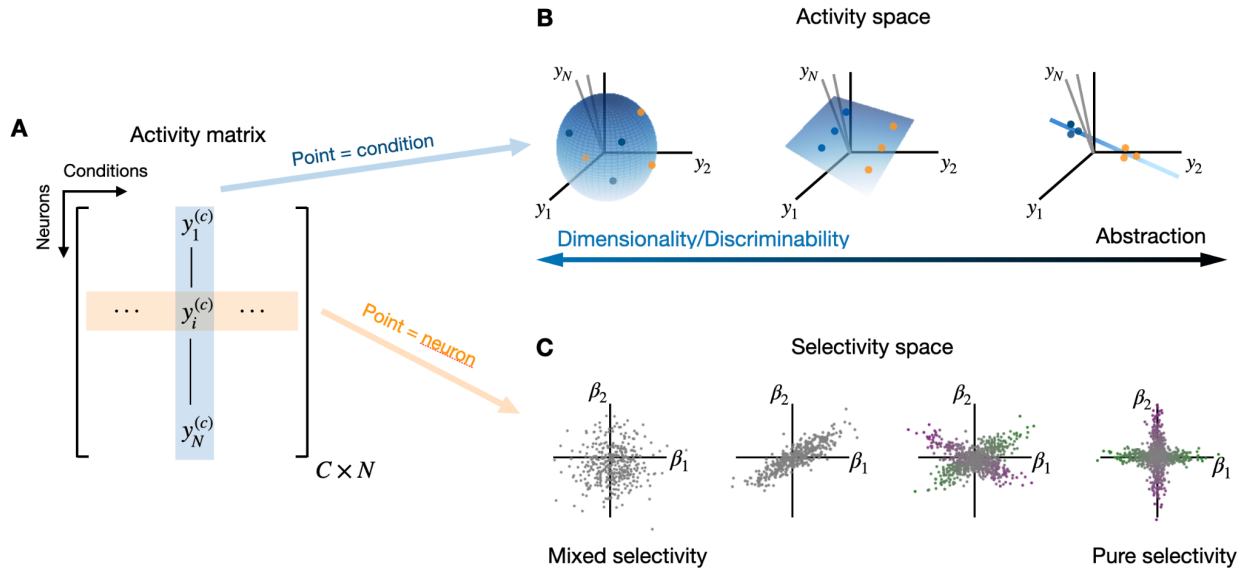

**Figure 2**: Characterizing structure in neural activity.
A. The activity of a population of N neurons across C trial conditions forms a CxN *activity matrix*, where each column is the population activity in one trial condition, and each row is the activity of one neuron across all trial conditions.
B. The structure of population activity can be described in terms of geometry in the activity state space, where each axis is the activity of one neuron. Each column of the activity matrix defines one point, which can be colored based on the behavioral response in that particular condition. The spatial arrangement of the points describes the neural representation and determines its computational properties, such *as flexibility, generalization, or abstraction [10].* Random patterns of activity (left) lead to a high-dimensional representation that allows for high discriminability and flexibility but low generalization. One-dimensional representations (right) instead maintain only information relevant to the output specific to the task, leading to high abstraction and generalization but low flexibility. In between these two extremes (middle), disentangled representations [10,71] enable high generalization while preserving information about several variables.
C. Each row of the activity matrix describes the response profile of an individual neuron and can be represented as a point in condition space (where each axis is the activity in one task condition, not shown), or in a lower-dimensional selectivity space where each axis is a measure of selectivity to a task variable (e.g. the linear regression coefficients). The resulting distribution of points can be used to assess modularity, defined here as the presence of clusters in the conditions or in the selectivity space. Unstructured mixed selectivity (left) corresponds to a single isotropic cloud of points. In classical pure selectivity (right), individual neurons instead form clusters aligned with individual task variables. This is also called a categorical representation [4]. In between these two extremes, neurons can form single or multiple groups with anisotropic mixed selectivity (middle two panels). Purple and green colors illustrate two sub-populations identified by a Gaussian-mixture clustering algorithm, with color shade indicating the probability of assignment to each cluster (gray indicates random assignment). See also [18].

## 3. Relating the structure of activity and connectivity using network models

Network models have become essential tools for understanding how specific computations may be related to the activity structure at different stages of processing. The transformation from one area to the next is typically modeled using a simple network model which receives inputs from an upstream area and is read out by downstream neurons (Fig. 3). Starting from hypotheses on the input structure, computational studies have used training algorithms to adjust connectivity weights and generate networks that perform specific tasks [27,74–77]. The activity in the network can then be examined with the same methods as for neural recordings and compared to them [76,78,79]. Beyond activity, such trained networks directly provide effective connectivity weights between neurons and therefore open up the possibility of examining an additional level of structure typically not accessible in experiments. Here we describe how the connectivity structure can be analyzed in a manner directly analogous to neural activity.

For concreteness, we consider a feed-forward network where an intermediate layer of N neurons receives inputs from M upstream units and sends outputs to K readouts (Fig. 3 A). The connectivity in this model consists of two parts: (i) inputs from upstream units to the intermediate layer, which form a vector of N weights for each of the M input units; (ii) readouts from the intermediate layer to downstream units, which form a vector of N weights for each of the K output units. The connectivity can therefore be represented as a Nx(M+K) *weight matrix* (Fig. 3B), on which one can perform the same analyses as on the activity matrix (Fig. 2). Indeed, each column is a vector over neurons, and each row contains weights received or sent out by an individual neuron in the intermediate layer. Importantly, this representation of connectivity in terms of a weight matrix is not specific to feed-forward models. In fact, it was first introduced in RNNs with low-rank connectivity structure [80,81], a broad class of models where the relation between connectivity, dynamics, and computations can be understood in detail [82–88]. Unrolling the temporal dynamics in such a network, the recurrent connectivity forms an extended set of input and output weights (Figure 4).

Each column of the weight matrix, defines a vector, or direction, within the N-dimensional activity state space of the intermediate layer (Fig. 3 C). Vectors corresponding to inputs and outputs play different roles. In a linear network, each input vector determines the direction in state space along which the activity in the intermediate layer varies when only one input is activated. Readout vectors instead specify the set of directions to which the outputs are sensitive, while directions orthogonal to them are output-null and therefore, "private" to the intermediate layer [89–91]. Altogether, the geometric arrangement between readout and input vectors fully specifies how a linear network transforms inputs into outputs.

Focusing on the rows of the weight matrix leads to a complementary view. Each row corresponds to one neuron in the intermediate layer and contains the set of input weights that this neuron receives and the set of readout weights it sends out. Each row therefore defines a point in the *connectivity space* where axes represents input and output weights (Fig. 3 D). A full network leads to a distribution of points in the connectivity space, one for each neuron.Different low-dimensional projections of that distribution provide complementary types of information. The distribution of input weights determines the selectivity to different inputs [81], while the distribution of output weights is related to choice probabilities [92]. More generally, the structure of the distribution in connectivity space can be analyzed using the same methods as when examining modularity in the selectivity space (Fig. 2C) to identify groups of neurons that share common patterns of weights. Each group is defined in the full connectivity space but can reflect correlations between input weights, input and output weights, or both input and input-output weights. Such analyses can reveal additional structure which is not always directly apparent in the activity or even in the geometry of connectivity and can uncover supplementary computational mechanisms that we further describe below.

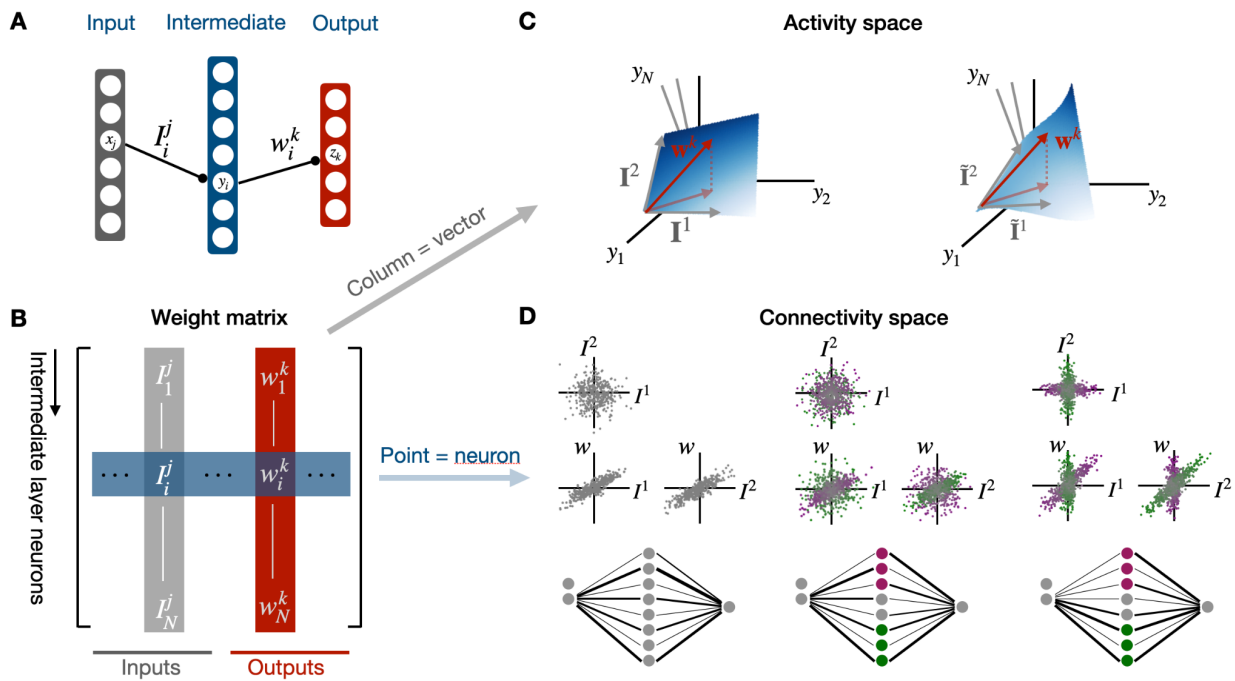

**Figure 3**: Characterizing connectivity structure.
A. Network model representing recorded neurons as an intermediate layer receiving inputs from an upstream area and read out by downstream output units. Each neuron *i* in the intermediate layer receives weights $I_i^j$ from input *j* and sends weights $w_i^k$ to readout *k*.

B. The full connectivity can be represented as a *weight matrix* where row *i* contains the weights received or sent out by neuron *i* in the intermediate layer, and each column contains all the weights sent out by an input unit or received by an output unit.

C. Each column of the weight matrix defines a vector in the activity state space of the intermediate layer. In a linear network (left), vectors $I^j$ of input weights determine the embedding of inputs in the activity space, while output vectors $w^k$ determine the directions being read out. In a non-linear network (right), the linear input manifold is bent by the non-linear activation function, and input vectors rescaled by the local gain determine local tangent planes.

D. Each row of the weight matrix can be represented as a point in *connectivity space* where each axis is the synaptic weight with respect to one input or output unit. Different projections of the resulting cloud of points contain different types of information, illustrated here for networks with two inputs and one output unit. The modularity in connectivity is defined by clusters in this connectivity space. Top row: Distributions of input weights directly determine the selectivity with respect to inputs, and the resulting modular structure in the activity. Middle row: correlations between input and output weights determine the relationship between each input and readout. Bottom row: illustration of corresponding networks. Colors indicate neurons belonging to different clusters. Left: A single cluster implies that all neurons have statistically identical mixed selectivity and share the same pattern of correlations between input and output weights. No structure appears in the neural activity in the middle layer (all neurons are grey). Middle column: Multiple clusters can appear based solely on correlations between input and output weights, implying that different sub-populations transfer different inputs to the readout unit despite having statistically identical mixed selectivity with respect to the two inputs. Right: Alternatively, clusters can be defined based on structure in both input weights and input-output correlations, corresponding to sub-populations with pure selectivity transferring different inputs.

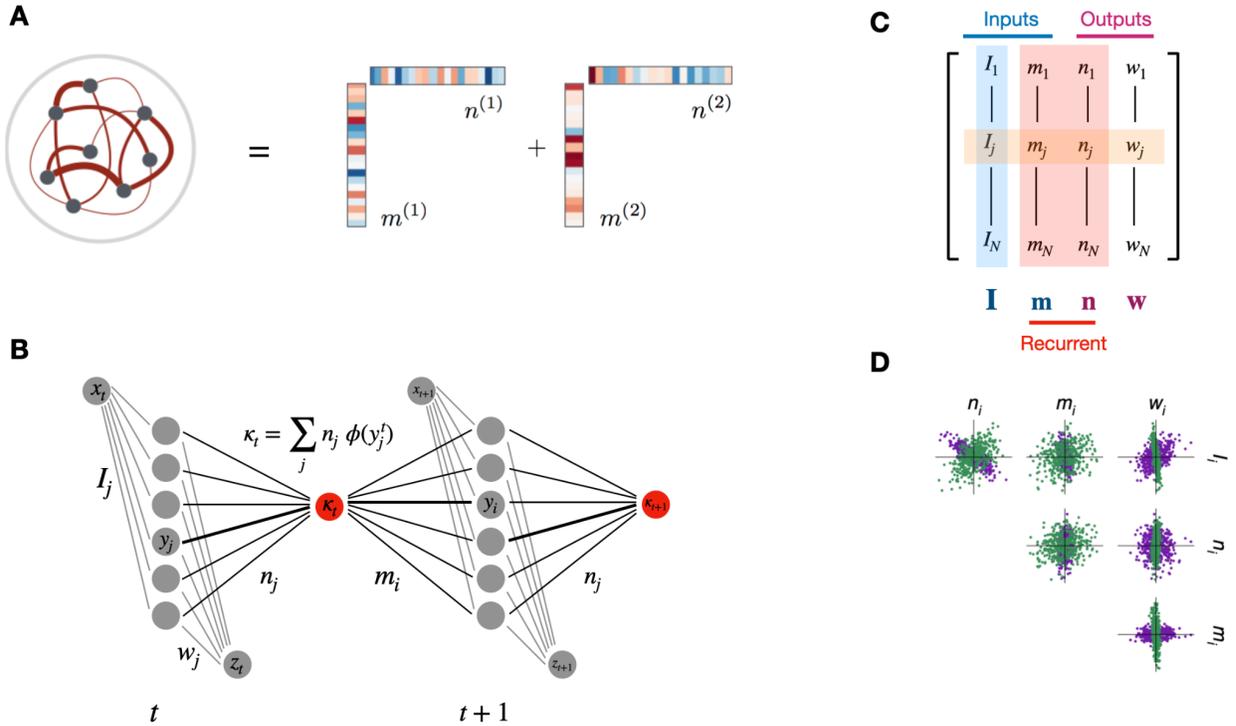

**Figure 4**: Characterizing connectivity in low-rank recurrent networks. A: In low-rank recurrent networks, the recurrent connectivity matrix can be represented as a sum of unit-rank terms consisting of pairs of column and row vectors $m^{(r)}$ and $n^{(r)}$. B: Unrolling temporal dynamics in discrete timesteps, each unit-rank term defines an effective feedback loop that integrates activity into a latent variable $\kappa_r$, which effectively reads out the activity at the previous time step through the row connectivity vector $n^{(r)}$ (a single unit-rank term is represented in this illustration). This latent variable is then fed back into the network at the next time step through weights determined by the column vector $m^{(r)}$. C: The connectivity in low-rank recurrent networks can described by including the column and row connectivity vectors into the weight matrix. The column vectors $m^{(r)}$ play the role of effective inputs, while row vectors $n^{(r)}$ form effective outputs. The geometry and modularity in the resulting weight matrix can be examined in the same manner as for feed-forward networks (Fig. 3). D: In particular, the modular structure can be assessed in terms of clusters in the connectivity space, where every line of the weight matrix is represented as a point [81].

## 4. Examples for specific types of computations

Having set up a broad framework for assessing geometric and modular structure in both neural activity and connectivity, we next apply it to review recent lines of work that examined three different classes of computations.

## 4.1 Flexible classification of random input patterns

A long tradition of theoretical works has focused on the classification or memorization of random patterns [64,66,93–97]. While in the real world, the inputs are typically structured (non-random), and often similar to each other, this framework can be applied to model experiments based on associative learning of arbitrary stimuli (Fig. 1a) [29,30,98]. One of the key theoretical questions has been how the structure of activity and connectivity in the intermediate layer can optimize the flexibility of the network by maximizing the number of possible binary readouts [66,99].

At the level of structure in the activity, a central theoretical result is that the number of possible classifications grows exponentially with the embedding dimension of represenations in activity state space [100], so that expanding dimensionality between the input and intermediate layer increases the number of possible classifications [101,102]. From the point of view of individual neurons, high embedding dimensionality can be directly related to strong and heterogeneous non-linear mixed selectivity, which, therefore, directly favors flexible classification of random inputs [2,103].

At the level of connectivity, high embedding dimensionality and non-linear mixed selectivity can simply be achieved by assigning random, unstructured connectivity weights between the input and intermediate layer. A series of works has examined the influence on dimensionality and classification of different features of this random input connectivity, such as sparsity [95,96] and degree of connectivity [97,104]. In this framework,
 learning specific classifications is therefore achieved by adjusting only the weights between the intermediate and the output layer. This situation is closely related to random feature models in machine learning [105–107], which lie at the heart of recent investigations of the neural tangent regime in deep networks [108–110]. Similarly, a large range of temporal inputs can be generated by adjusting the weights of readouts from randomly connected recurrent neural networks [48,111–113].

Altogether, when input patterns are unstructured, highly flexible outputs can be achieved with a fully random structure in both activity and connectivity, implying a high-dimensional embedding geometry and a lack of modularity as defined in Fig. 2. An important challenge to unstructured networks is however their limited ability to generalize to previously unseen inputs, as increasing dimensionality in the intermediate layer through non-linear random projections may potentially separate similar patterns of activations in the input layer. Recent works have therefore considered more structured inputs and outputs.

## 4.2 Structured inputs and readouts

When faced with naturalistic stimuli, humans and other animals have the capacity to infer correct responses to previously unseen inputs. This ability to generalize is hypothesized to rest on an inherent structure of the physical and social world [114]. Indeed, although naturalistic inputs are high-dimensional in terms of the patterns of activations of sensory receptors and neural responses in early sensory areas [115], they are formed by physical and social objects that are, in general, lower dimensional. The *manifold hypothesis* therefore states that naturalistic stimuli can be modeled in terms of manifolds of relatively low intrinsic dimension, embedded non-linearly in a much higher-dimensional space representing sensory activations such as patterns of photo-receptors on the retina [116,117]. This hypothesis is in fact, implicit in classical categorization tasks, where the experimenter varies a few task variables defining the intrinsic dimension of the input manifold, but individual sensory stimuli are high dimensional (Fig 1 B,C).

Recent works have sought to incorporate the manifold hypothesis in the framework of network models by assuming that the set of patterns in the input layer is generated from a distribution based on a hidden low-dimensional manifold embedded in a high-dimensional space [67,118–120]. The desired responses in the output layer are then determined by the intrinsic latent variables of the manifold rather than the high-dimensional inputs themselves [67]. Such an input-output structure clearly allows for generalization by interpolation on the hidden manifold, yet this does not guarantee that a network trained on a task necessarily achieves generalization [67,121,122]. The central question then becomes what type of structure in the network's activity and connectivity best matches the hidden manifold structure, in the sense of enabling the network to generalize.

A key theoretical proposal is that generalization is optimized when the geometry of the activity in the intermediate layer represents the input manifold linearly by minimizing the embedding dimensionality [10]. Such representations are referred to as *abstract* because they allow for generalization as in cases in which all the irrelevant information is discarded, and only one abstract variable is encoded (i.e. when the representation is disassociated from specific instances, which is a defining characteristic of abstraction). They are alternatively known as factorized [123] or disentangled representations [124,125], as independent task variables are represented along orthogonal axes in the activity state space. In these representations, the coding directions of each abstract variable are approximately parallel and hence enable linear readouts to directly generalize across values of irrelevant variables (Fig. 2B).
Recent computational work has shown that a network model such as in Fig 3 directly acquires an abstract, disentangled representation in the intermediate layer when trained to perform multiple classifications on high-dimensional inputs generated from an underlying hidden manifold [67]. Analogous disentangled geometries have been found in recurrent neural networks optimized to generalize in temporal tasks such as working memory [126] or flexible timing [122]. Signatures of abstract representations have also been identified in experimental recordings in a variety of brain areas, including the monkey face-representation areas [71,127,128], the DMPFC when animals perform flexible timing [122,129–131], the somatosensory cortex[132], or the prefrontal cortex and hippocampus when animals performed a task based on an abstract structure [10].
The observed representations are not perfectly disentangled, but display for high generalization and flexibility.

## 4.3 Context-dependent readouts

In naturalistic conditions, a given stimulus often requires different responses depending on the overall situation in which it occurs. A number of experimental studies have examined the neural bases of such flexible behavior by presenting identical stimuli within different contexts that are either explicitly indicated [3,44,133,134] or implicitly inferred [45,135–137].

A paradigmatic example is context-dependent perceptual decision-making, where stimuli consist of superpositions of two features, such as motion and color [3]. Depending on a contextual cue, the goal of the task is to integrate either one or the other (Fig. 1 D). Analyses of neural activity recorded in the monkey prefrontal cortex during this task have identified highly mixed selectivity that lacked any apparent modular structure, and focused therefore on the geometry of population activity [3,138]. Studies of trained network models, both recurrent [3,81,87,134,139,140] and feed-forward [133] have examined the mechanisms underlying this task. While all models reproduced the structure in the geometry of activity, some models led to additional modular structure in selectivity [133,139], while others argued for a modular structure in connectivity, but not necessarily selectivity [81].

What type of structure is then strictly needed for context-dependent decision-making, and what type of structure is a byproduct of specific modeling choices? This question is most easily addressed in single-layer feed-forward models with a simple threshold-linear transfer function [133,141]. Assuming the inputs to the network are factorized along the two stimulus features (Fig. 5), the goal of the output is to reproduce one or the other feature depending on the contextual cue. Ref. [133] showed that training networks on this task can lead to two types of solutions depending on the initialization of the connectivity weights. If output weights are initialized to strong values, the network is in the so-called "lazy" or "neural tangent kernel" regime, where only the output weights are effectively trained [108–110,142]. As a result, the input weights remain at their initial random values, and the selectivity to stimuli in the intermediate layer is fully random, implying that neural activity lacks any modular structure. A modular structure is however present in the connectivity at the level of correlations between input and output weights (Fig. 5). If output weights are initialized with lower values, the network is instead in the so-called rich regime, where both output and input weights are modified during training [143–146]. As a consequence, additional structure develops in the input weights and leads to a modular structure in the selectivity of the intermediate layer, while the modular structure in correlations between input and output weights is still present. Altogether, whether the modular structure is apparent in the *activity* therefore depends on the details of learning parameters. The structure in *connectivity,* based on the correlations between input and output weights, is instead a fundamental constraint for implementing the computation. While these insights are most transparently reached in simplified feed-forward networks, analogous results have been obtained in recurrent models with low-rank connectivity [81,87]. In these networks, the recurrent connectivity can be factored into sets of effective input and effective output weights (Fig. 4). The correlations between effective output weights and external inputs then define a modular connectivity structure, in which two sub-populations of neurons integrate the two input features separately. In a manner similar to feed-forward networks, this connectivity structure may lead to modular structure in selectivity but does not need to, depending on the details of network training [81].

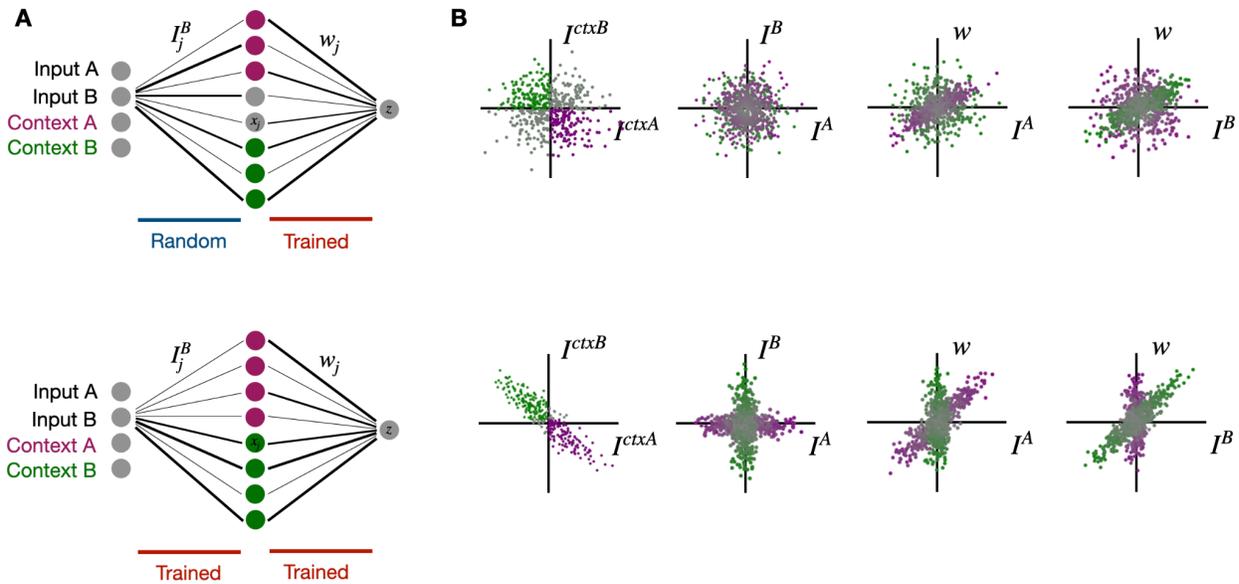

**Figure 5**: Connectivity structure in context-dependent decision making. Illustration of results in feed-forward networks trained in two different regimes (based on [133,141]).
**A**: illustration of the network. The intermediate layer receives two stimulus inputs A and B and two binary contextual inputs. In context A, the network needs to output the value of stimulus A, and conversely, in context B.
**B:** Distribution of connectivity weights as in Fig 3D.
The neurons in the intermediate layer are split into a task-irrelevant population (gray) and two task relevant populations (purple and green). Assuming a threshold-linear transfer function, based on contextual inputs (B leftmost panel), purple neurons are active in context A but not context B, and conversely for green neurons. For purple neurons weights of input A are correlated with the readout weights, ensuring that when active this population transmits stimulus A to the output. The converse holds for the green population. Top: In the lazy learning regime, only readout weights are modified during learning. The purple and green populations are therefore mixed selective to the two stimuli, and the modular structure is based on the correlation between input and output weights combined with the sign of the contextual inputs. Bottom: in the rich learning regime, all weights are trained, so that the task-irrelevant population shrinks and the two task-relevant populations acquire pure selectivity to potentially both inputs and context. In that situation, a modular structure is present also in the selectivity of the neurons.

**Discussion**
Diversity is a prominent feature of the brain: neurons are morphologically, genetically, and functionally diverse, and each of them is connected to a different subset of other cells. So it is unsurprising that different neurons typically respond differently to the same sensory input. However, these responses are not completely disorganized, and it is often possible to find interesting structures which reflect reproducible and interpretable patterns in the statistics of the responses of populations of neurons. Here we reviewed recent approaches to identify, study and interpret these structures. Interestingly, the approach used to reveal structures in neural activity can also be used to study structures in afferent and efferent connectivity patterns. For both the neural activity and the connectivity, one can study the geometry of the neural representations, which is directly related to the computational properties relevant to a linear readout. A complementary analysis of the same neural data can reveal some form of organization at the level of the responses of individual neurons. Modular representations, also called categorical [4,18,21], are observed when the responses of individual neurons to multiple experimental conditions are not completely unstructured, as for example, when groups of neurons tend to respond in a similar way to different sensory stimuli. Modularity could be the consequence of anatomical organization at different length scales It is very clearly observed at a large spatial scale in the brain: all neurons in the visual cortex tend to respond more strongly to visual stimuli than the neurons in the auditory cortex. It could also reflect the existence of different types of neural cells: for example, inhibitory and excitatory cells often have different response properties.

More recent studies have clearly demonstrated that highly diverse and structured neural responses reflect the diversity of genetic profiles of individual cells [147].

However, modularity could also result from a learning process and reveal itself at much smaller spatial scales, e.g. within a single brain region or a single cortical column. Finally, as we discussed, modularity might be detected in the connectivity but not necessarily in the patterns of neural activity, though typically, the two structures are related.

What are the computational implications of modularity? For the representational geometry, it is possible to say something when one assumes that the readout is linear, and in some situations under this assumption, it is possible to predict the behavior of the subject [2]. For modularity, it is more challenging because the diversity of the responses to different experimental conditions is not read out directly by downstream neurons, and, at the same time, we know very little about the relation between modularity and representational geometry. The recent developments in recording techniques, which now offer the possibility of knowing much more about the type of recorded neuron and its connectivity, will enable us to reveal many important structures, and it is essential that we start studying now their possible computational implications.

**Outstanding Questions**
- How are the functional properties of neurons during a task related to their biological properties such as gene expression, physiology, and connectivity? Emerging recording techniques allow experimentalists to collect both functional and biological information for the same set of neurons. This opens the possibility to relate biological labels with functional labels as obtained for instance from analyses of modularity, and is likely to reveal new levels of structure.
- Trained artificial networks have become an important model system for studying the relation between structure and function in fully-observable systems of neuron-like elements. Training algorithms used for such models do not aim for biological plausibility but offer an efficient tool to explore the space of solutions for a given computation. It remains to understand to which extent the resulting network structure reflects general computational constraints rather than the peculiarities of the training algorithm or initial conditions. It is therefore important to further study different learning regimes in artificial networks and in particular recurrent ones [148].
- Achieving a theory of the relation between structure and function in the brain ultimately requires having a map of the space of computations underlying naturalistic behavior. The simplistic characterization attempted here (Fig 1) is based on laboratory tasks originating from the psychology literature. The underlying taxonomy of cognitive functions has been recognized as largely ambiguous and in need of reassessment [149].
- Here, we have focused on the potential roles of different types of structures in computations. The structure of the brain however clearly reflects other constraints, and in particular, the fact that biological networks are generated through developmental dynamics. Models combining computational, developmental, and other types of constraints will be essential for understanding the structure of the brain.

**Acknowledgments**
SO thanks the Princeton Neuroscience Institute for its hospitality during the writing of this piece. SO was supported by the CRCNS project PIND (ANR-19-NEUC-0001-01), the NIH Brain Initiative project U01-NS122123, the program "Ecoles Universitaires de Recherche" launched by the French Government and implemented by the ANR, with the reference ANR-17-EURE-0017 and a CV Starr Fellowship from the Princeton Neuroscience Institute. SF was supported by the Neuronex NSF grant, the Simons Foundation, the Gatsby Charitable Foundation and the Swartz Foundation.